\documentstyle[11pt,newpasp,twoside,epsf]{article}
\def\edcomment#1{\iffalse\marginpar{\raggedright\sl#1\/}\else\relax\fi}
\reversemarginpar

\begin{document}
\title{NEW DETREMINATION METHOD OF PRIMORDIAL Li ABUNDANCE}
\author{T. Kajino, T.-K. Suzuki, S. Kawanomoto, and H. Ando}
\affil{National Astronomical Observatory \\ 
The Graduate University for Advanced Studies \\
Mitaka, Tokyo 181-8588, Japan\\
Department of Astronomy, University of Tokyo \\
Bunkyo-ku, Tokyo 113-0033, Japan}

\begin{abstract}
We discuss the primordial nucleosynthesis in 
lepton asymmetric Universe models.  In order to better estimate 
the universal baryon-mass density parameter $\Omega_b$, 
we try to remove the uncertainty from the theoretical prediction of
primordial $^7$Li abundance.  We propose a new method to
determine the primordial $^7$Li by the use of isotopic abundance
ratio $^7$Li/$^6$Li in the
interstellar medium which exhibits the minimum effects
of the stellar processes.
\end{abstract}

\section{Introduction}

Recent spectral and photometric observations of 
Type Ia supernovae at high redshifts 
(Riess et al. 1998; Perlmutter et al. 1999) have raised 
a possibility that the cosmic expansion is accelerated.
For a flat cosmology these data have $\chi^2$-minimum around
$\Omega_0 \approx 0.3$ and $\Omega_{\Lambda} \approx 0.7$,
allowing Hubble time $\sim$15Gyr which is not inconsistent with
the age of the Milky Way constrained from the observations
of the oldest globular clusters.

Cosmological model for primordial nucleosynthesis provides 
independent method to determine $\Omega_0$.
The Big-Bang nucleosynthesis model (Copi et al.~1995)
predicts $0.04 \le \Omega_b h_{50}^2 \le 0.08$.  
Combining this value with X-ray observations of rich
clusters that indicate 0.3$h_{50}^{-3/2} \approx \Omega_b$/$\Omega_0$ 
(Bahcall et al.~1995; White et al.~1993),
total $\Omega_0$ turns out to be 
$\Omega_0 h_{50}^{1/2} \approx$ 0.1$\sim$0.3,
which is consistent with flat cosmology.

However, in the determination of $\Omega_b$, 
a difficulty has been imposed by 
recent detections of a low deuterium abundance,
2.9$\times$10$^{-5}$ $\le$ D/H $\le$4.0$\times$10$^{-5}$,
in Lyman-$\alpha$ clouds along the line of sight 
to high red-shift quasars (Burles \& Tytler 1998ab).
Primordial abundance of $^7$Li is 
constrained from the observed "Spite plateau", 
0.91$\times$10$^{-10}$ $\le$ $^7$Li/H$\le$1.91$\times$10$^{-10}$
(Ryan et al. 2000a), 
and the $^4$He abundance by mass, 0.226$\le Y_p \le$0.247
(Olive et al. 1999),
from the observations in the HII regions.
In order to satisfy these abundance constraints 
by single $\Omega_b$ value, one has to assume an appreciable 
depletion in the observed abundance of $^7$Li,
which is still controversial both theoretically and observationally.
We are now forced to critically study the uncertainty.
An independent method to determine the primordial 
$^7$Li is also desirable. 

\section{Primordial Nucleosynthesis}

\subsection{$^4$He vs. D and Neutrino Degeneracy}

Shown in Fig.~1 is the comparison between the observed abundance
constraints on $^4$He, D/H, and $^7$Li/H and the calculated curves
in the homogeneous Big-Bang model as a function of $\eta$,
where $\eta = n_B/n_\gamma$ and 
$\Omega_b h^2_{50}$ = $\eta$ 1.464$\times$10$^8$.  
Solid curves display the theoretical prediction of
primordial abundances in the standard particle model for neutrino,
which preserves the lepton symmetry $L_{\nu}$=0.

\begin{figure}[ht]
\begin{center}
\epsfxsize=8cm
\epsfysize=12cm
\mbox{   }\epsfbox{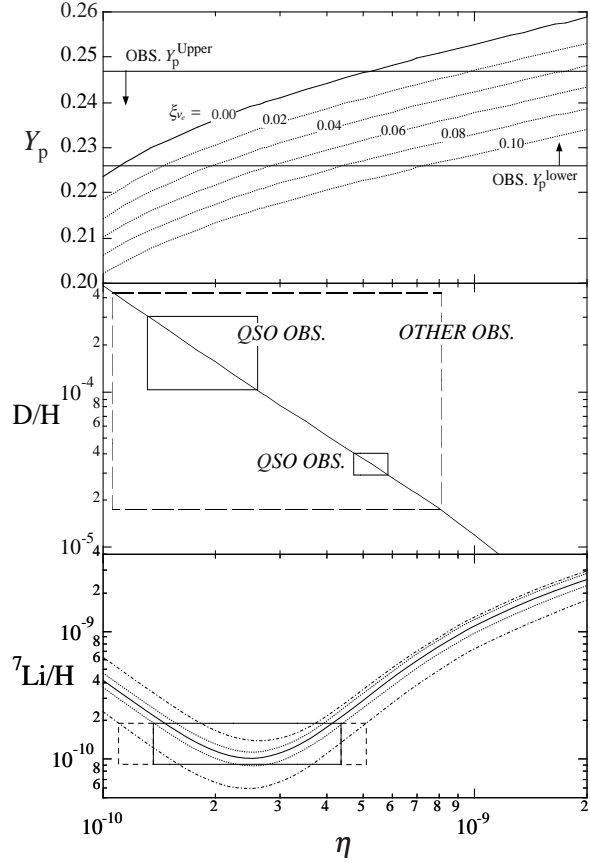}
\caption{$Y_p$, D/H and $^7$Li/H vs. $\eta$ for various 
neutrino degeneracy parameters $\xi_{\nu_e}$.
Observed constraints on D/H are from Rugers \& Hogan (1996) (upper) 
and Burles \& Tytler (1998ab) (lower).
Theoretical curves for $^7$Li/H show 
the 2$\sigma$ uncertainties based on two
different error estimates of the reaction rates.
See text for details.}
\end{center}
\end{figure}

There is now a good collection of abundance
information on the $^4$He mass fraction, $Y_p$, in over 50
extragalactic HII regions, from which the upper limit 
on primordial abundance, $Y_p \le 0.240$, and a systematic error, 
$\Delta Y_{sys}$=0.005, were extracted.  Unfortunately, 
for this upper limit one cannot find $\Omega_b$ to satisfy 
both abundance constraints on $^4$He and D/H.  
(See the solid curves in Fig.~1.)

It has been recognized that $\Delta Y_{sys}$ may even be larger
(Izatov et al. 1994; Thuan 2000), making
the upper limit as large as $Y_p \le$0.247.
If this upper limit is adopted, the Universe model
with $\eta \approx$5$\times$10$^{-10}$ is marginally consistent
with both abundance constraints.
However, since even smaller value, $Y_p$=0.235$\pm$0.003,
in low-metallicity extragalactic HII regions 
has been reported by Peimbert \& Peimbert (2000),
this potential conflict is to be studied more carefully.

One possible solution is to introduce a lepton asymmetry.
Theoretically, it is natural to assume that both baryon
and lepton symmetries are simultaneously broken, 
$B\neq$0 and $L_{\nu} \neq$0, 
due to the CP violation in baryogenesis.
$L_{\nu} \neq$0 is fulfilled by 
neutrino degeneracy with non-zero $\xi_{\nu_e}$,
where $\xi_{\nu_e} = \mu_{\nu_e} / kT_{\nu}$ and
$\mu_{\nu_e}$ is the chemical potential of electron neutrino.
Since neutrinos had energy density comparable to the densities 
due to photons and charged leptons in the early Universe,
even a small degeneracy 0$<\xi_{\nu_e}\ll$1 leads to 
an appreciable decrease in the neutron-to-proton number ratio,
slightly faster acceleration of the Universal expansion, and
a small increase of the weak-decoupling temperature.
As a net result, $^4$He abundance decreases 
with increasing $\xi_{\nu_e}$,
as shown in Fig.~1, while keeping D/H and $^7$Li/H 
almost unchanged in logarithmic scale (Kajino \& Orito 1998).
Since the abundance constraint on primordial $^4$He is more
accurate than the other light elements, this helps determine
the most likely $\xi_{\nu_e}$.
$\xi_{\nu_e} \sim 0.05$ can best fit the $^4$He abundance as well
as low deuterium abundance D/H$\sim$10$^{-5}$,
leaving inevitable requirement that the observed abundance level
of Spite plateau, $^7$Li/H$\sim$10$^{-10}$, should be
the result of depleted primordial abundance.

\subsection{$^7$Li vs. D}

There are several input parameters in the primordial 
nucleosynthesis calculation.  
As the number of light neutrino families $N_{\nu}$ = 3 and 
the neutron lifetime $\tau_n$ = 886.7 $\pm$ 1.9 s are known, 
the remaining major uncertainty arises from
input nuclear reaction data.
We did not take account of the effects of sterile neutrino
which is a hypothetical particle for interpreting flavor mixing.

Laboratory cross section measurements ever done 
provide rather precise thermonuclear reaction rates for the 
production of D, T, $^3$He, and $^4$He.
It however was claimed in literature (Smith et al. 1993)
that the $^7$Li abundance is strongly subject to
large error bars associated with the measured cross sections
for $^4$He($^3$H,$\gamma$)$^7$Li at $\eta \la$ 2$\times 10^{-10}$ 
and $^4$He($^3$He,$\gamma$)$^7$Be at 3$\times 10^{-10} \la \eta$.
There are in fact several inconsistent data with one another,
leading to large uncertainty in the primordial $^7$Li, 
as displayed by long-dash-dotted curves in Fig.~1. 

We studied these two reactions very carefully 
and concluded that the proper 2$\sigma$ error bars could be 
1/4$\sim$1/3 of the previous ones (Kajino et al. 2000). 
This improvement owes mostly to, first, the new precise measurement
(Brune et al. 1994) of the cross sections for $^4$He($^3$H,$\gamma$)$^7$Li
and, second, the systematic theoretical studies 
of both reaction dynamics and quantum nuclear structures of
$^7$Li and $^7$Be, whose validity is critically
tested by electromagnetic form factors measured by
high-energy electron scattering experiments.

When our recommended error estimate is applied to the determination
of $\Omega_b$ in Fig.~1, we lose $\Omega_b$ value to explain both D/H
and $^7$Li/H simultaneously.
If we allow for larger primordial $^7$Li abundance
in Population II halo stars because

\begin{figure}[ht]
\setlength{\unitlength}{1cm}
\begin{picture}(10,8)(-5,-1)
\epsfxsize=11cm
\put(-4,-1){\epsfbox{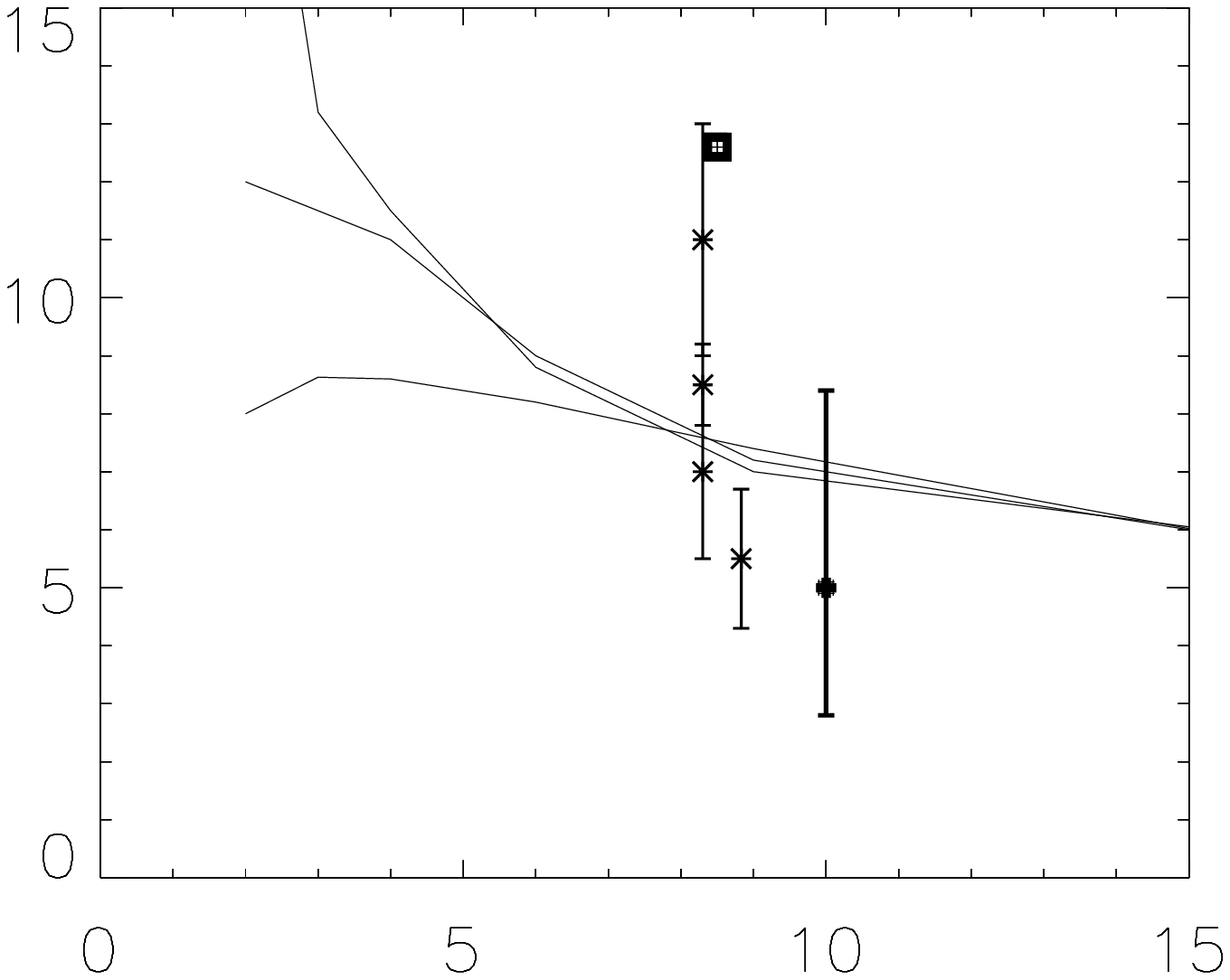}}
\put(3.3,5.8){\small solar-meteorite}
\put(4,2.5){\small ${\chi}^2$-Ori (Kawanomoto 2000)}
\put(3.5,3){\small $\zeta$-Per Meyer (1993)}
\put(3.3,3.5){\small $\zeta$-Oph Meyer (1993)}
\put(3.3,4.1){\small $\zeta$-Oph Leimoine (1995)}
\put(3.3,5){\small $\rho$-Oph Ferlet (1983)}
\put(-3.75,3.5){\large \bf $^7$Li/$^6$Li}
\put(1.5,-0.7){\large \bf R(kpc)}  
\put(-1,6){\small 2 Gyr}
\put(-1,5){\small 6 Gyr}
\put(-1,4.3){\small 12 Gyr}
\end{picture}
\caption{Calculated isotopic abundance ratios $^7$Li/$^6$Li at 
various times as a function of the Galactocentric distance R,
compared with observed data.}
\end{figure}

\noindent{of possible lithium
depletion for diffusion or rotation-induced mixing of matter
(Deliyannis et al. 1998 ; Pinsonneault et al. 1992)
or some systematic uncertainty in the model atmospheres (Kurucz 1995),
we can recover the concordance.  Taking depletion factor 
$\approx$ 2.5, $\Omega_b h_{50}^2 \approx$ 0.075 
best fits all abundance constraints in the homogeneous Big-Bang model.
Note that larger $\Omega_b h_{50}^2 \approx$ 0.2 is allowed 
in the inhomogeoeus Big-Bang model (Kajino \& Orito 1998).}

\section{$^7$Li/$^6$Li Ratio in the Interstellar Medium (ISM)}

The lithium in the ISM is almost free 
from the complicated stellar processes.
A diffuse cloud along the line of sight to $\zeta$Oph
was observed to show the lithium abundance depleted by 1.58dex 
from the meteoritic solar-system value 12.3.  
This is due to dust grain formation (Savage \& Sembach 1996).  
The isotopic ratio is free from such condensation effects and 
represents the real ratio of chemical compositions in the gas phase.
The D/H (Wannier 1980) and $^3$He/H (Rood et al. 1995) abundance 
ratios in the ISMs have been observed over wide Galactocentric 
distance range 0$\le$R$\le$12kpc and used to constrain 
the primordial abundance of D/H (Dearborn et al. 1996), but
the distribution of $^7$Li/$^6$Li was poorly known.  

\subsection{Observation}

Observations of isotopic abundance ratio, $^7$Li/$^6$Li,
have been performed by several groups
(Ferlet \& Dennefeld 1983, Lemoine et al. 1993, 1995, Meyer et al. 1993)
only for the ISMs in our solar neighborhood.
The observed ratio is less than 12.3 and larger than 2.1 
which is a predicted GCR abundance ratio.

\begin{figure}[ht]
\setlength{\unitlength}{1cm}
\begin{picture}(10,8)(-5,-1)
\epsfxsize=11cm
\put(-4,-1){\epsfbox{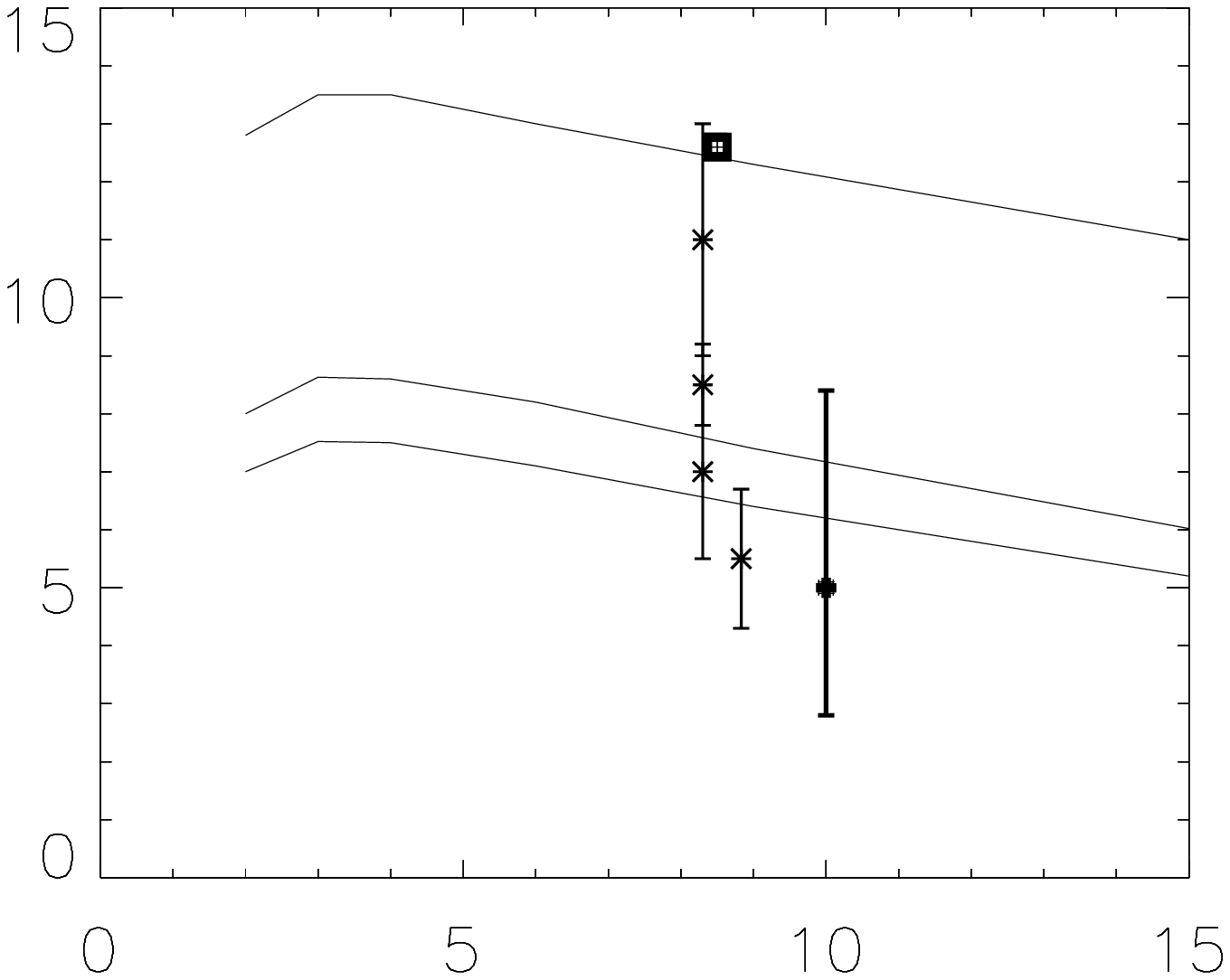}}
\put(3.3,5.8){\small solar-meteorite}
\put(4,2.5){\small ${\chi}^2$-Ori (Kawanomoto 2000)}
\put(3.5,3){\small $\zeta$-Per Meyer (1993)}
\put(3.3,3.5){\small $\zeta$-Oph Meyer (1993)}
\put(3.3,4.1){\small $\zeta$-Oph Leimoine (1995)}
\put(3.3,5){\small $\rho$-Oph Ferlet (1983)}
\put(-3.75,3.5){\large \bf $^7$Li/$^6$Li}
\put(1.5,-0.7){\large \bf R(kpc)}
\put(-0.7,5.5){\small $^7$Li$_p=1.5\times 10^{-9}$}
\put(-0.7,4.3){\small $3.5\times 10^{-10}$}
\put(-0.7,3.1){\small $1.4\times 10^{-10}$}
\end{picture}
\caption{Calculated isotopic abundance ratios $^7$Li/$^6$Li at
the present time, $t_G$ = 12Gyr, for various primordial abundances,
$^7$Li$_p \equiv ^7$Li/H(primordial), as a function of the 
Galactocentric distance R, compared with observed data.}
\end{figure}

Using the Coude spectrograph of the 74-inch telescope
at Okayama Astrophysical Observatory, Japan, we have
succeeded for the first time in the determination of $^7$Li/$^6$Li
in the diffuse cloud along the line of sight to $\chi^2$Ori, which is
a member of OB association Gem-OB1, being located
at R = 10kpc (Kawanomoto et al. 2000).
The telescope performance was R=43,000 (with slit width of 100 $\mu$m),
exp=50hours, and S/N=2,800.

We found a decreasing gradient of $^7$Li/$^6$Li,
as shown in Figs.~2 \& 3.
It is interpreted as a result of gradual extinction
of the stellar production of $^7$Li.

\subsection{A New Method to Determine Primordial $^7$Li}

In order to study the sensitivity to the primordial
lithium abundance, we have calculated Galactic chemical 
evolution (GCE) of lithium (Kawanomoto 2000).
We adopted a hybrid model (Ryan et al. 2000b) of the inhomogeneous
GCE model (Suzuki et al. 1999), which was constructed 
for the early evolution of metal-deficient stars, 
being smoothly connected with a simple one-zone GCE model 
for later evolution.  Five different sources of 
lithium production are included in this model:
Primordial nucleosynthesis, GCR interactions with ISM,
$\nu$-induced nucleosynthesis in Type II SNe,
AGB star nucleosynthesis, and nova nucleosynthesis.
We took an approximation that each ring having different 
Galactocentric distance evolves independently so that
the observed present day star-formation-rate and the gas fraction 
are reproduced very well.
 
The calculated time variation of $^7$Li/$^6$Li is shown 
as a function of R in Fig.~2.
Remarkable decrease of $^7$Li/$^6$Li in the inner region
is caused by faster gas consumption for the star formation.  
It is discussed in literature that the meteoretic chemical
compositions are peculier and different from
those of ISM because they were possibly polluted by
nearby AGB star.  One might speculate another possibility
that the solar-system might ever have moved outward
over hundreds of turns of the Galactic disc, keeping high
$^7$Li/$^6$Li = 12.3 as it was in the original position 
when the solar system was isolated 
from viscous gas component at $t_G \approx 2\sim6$Gyr.

Figure~3 displays sensitivity of the $^7$Li/$^6$Li ratio 
at the present time $t_G$=12Gyr
to the primordial abundance of $^7$Li.
It is very sensitive to $^7$Li$_p$.
Except for old data point at $\rho$-Oph (Ferlet \& Dennefeld 1983),
which has the largest error bar among all data for
the solar neighborhood,
the observed ratios look more consistent with 
$^7$Li$_p$ = (1.4$\sim$3.5)$\times$10$^{-10}$
than $^7$Li$_p$ = 1.5$\times$10$^{-9}$. 
More data with smaller error bars are highly desirable
in order to convince the gradient of the $^7$Li/$^6$Li ratio
and to determine the primordial abundance of $^7$Li in this method.

%



\begin{references}
Bahcall, N.A., Lubin, L.M., \& Dorman, V. 1995, ApJ 447, L81.\\ 
Brune, C.R., Kavanagh, R.W., \& Rolfs, C. 1994, PR C50, 2205.\\
Burles, S., \& Tytler, D. 1998a, ApJ 499, 699; 1998b, ApJ 507, 732.\\
Copi, C.J., Schramm, D.N., \& Turner, M.S. 1995, ApJ 455, 95.\\
Dearborn, D.S.P. ,Steigman, G., \& Tosi, M. 1996, ApJ 465, 887.\\
Deliyannis, P., et al. 1998, ApJ 498, L147.\\  
Ferlet, R., \& Dennefeld, M. 1983, ApJ 409, L61.\\
Izatov, Y.I., Thuan, T.X., \& Lipovetsky, V.A. 1994, ApJ 435, 647.\\
Kajino, T., \& Orito, M. 1998, Nucl. Phys. A629, 538.\\
Kajino, T., Orito, M., Sakai, K., \& Deliyannis, P.C. 2000, in preparation.\\  
Kawanomoto, S., Ando, H., Kajino, T., \& Suzuki, T.-K. 2000, in preparation.\\
Kurucz, R.L. 1995, ApJ 452, 102.\\
Lemoine, M., et al. 1993, A\&A 269, 469; 1995, A\&A 298, 879.\\
Meyer, D.M., Hawkins,I., \& Wright, E.L. 1993, ApJ 409, L61.\\
Olive, K., Steigman, G., \& Walker, T. 1999, Phys. Rep., in press.\\
Peimbert, M., \& Peimbert, A. 2000, astro-ph/0002120.\\
Perlmutter, S., et al. (Supernova Cosmology Project Team) 1999, ApJ 517, 565.\\
Pinsonneault, M.H., Deliyannis, C.P., \& Demarque, P. 1992, ApJS 78, 179.\\
Rugers, M., \& Hogan, C.J. 1996, ApJ 459, L1.\\
Riess, A., et al. (High-z Supernova Search Team) 1998, AJ 116, 1009.\\
Rood, R. et al. 1995, Light Element Abundances, (ed. P.Crane, Springer) 201.\\
Ryan, S., Beers, T., Olive, K., Fields, B., \& Norris, J. 2000a,
 ApJ 530, L57.\\
Ryan, S.G., Kajino, T., Beers, T.C., Suzuki, T.-K., Romano, D.,
 Matteucci, F., \& Rosolankova, K. 2000b, ApJ, submitted.\\
Savage, B.D., \& Sembach, K.R. 1996, ARA\&A 34, 279.\\
Smith, M.S., Kawano, L.H., \& Malaney, R.A. 1993, ApJS 85, 219.\\
Suzuki, T.-K., Yoshii, Y., \& Kajino, T. 1999, ApJ 522, L125.\\
Thuan, T.X. 2000, in this volume.\\
Wannier, P.G. 1980, ARA\&A 18, 366.\\
White, S.D.M., et al. 1993, Nature 366, 429.\\
\end{references}
\end{document}